\documentclass[nofootinbib,aps,amssymb,floatfix,superscriptaddress,preprintnumbers]{revtex4}

\usepackage{graphicx}
\usepackage{bm}
\usepackage{amsmath}
\usepackage{amssymb}
\usepackage{amsfonts}
\usepackage{float}
\usepackage{hyperref}
\usepackage{dsfont}  
\usepackage{slashed}  
\usepackage{booktabs}
\usepackage{multirow}
\usepackage{subfigure}
\usepackage[sort&compress]{natbib}
\usepackage{xcolor}
\usepackage{ulem}

\newcommand{\be}{\begin{equation}}  
\newcommand{\ee}{\end{equation}}  
\newcommand{\beq}{\begin{eqnarray}} 
\newcommand{\eeq}{\end{eqnarray}}

\newcommand{\bea}{\begin{eqnarray}}
\newcommand{\eea}{\end{eqnarray}}

\newcommand{\nn}{\nonumber \\}

\begin{document}

\title{
Accessing the  gravitational form factors of the nucleon and nuclei \\ through  a massive graviton
}

\author{Yoshitaka Hatta}
\email{yhatta@bnl.gov}
\affiliation{Physics Department, Brookhaven National Laboratory, Upton, NY 11973, USA}
\affiliation{RIKEN BNL Research Center, Brookhaven National Laboratory, Upton, NY 11973, USA}

\begin{abstract}

In contrast to the electromagnetic form factors of the nucleon and nuclei that have been extensively studied in electron scattering, there is no known way to directly measure the gravitational form factors (GFFs), the off-forward hadronic matrix element of the QCD energy-momentum tensor.  I suggest exploring the possibility to  access the GFFs of the proton and nuclei in conjunction with massive graviton searches at future TeV-scale lepton-ion colliders. 

\end{abstract}

\maketitle

The gravitational form factors (GFFs) were 
originally introduced in the 1960s in order to study the mechanical properties of hadrons \cite{Kobzarev:1962wt,Pagels:1966zza,Polyakov:2018zvc} in much the same way as the electromagnetic form factors can reveal the distribution of the electric charge inside the nucleon and nuclei. However, despite their fundamental importance in the science of the hadron structure,  historically the GFFs have received far less attention than they deserve, in sheer contrast to the electromagnetic counterparts which  have been extensively studied  over the past 70 years \cite{Pacetti:2014jai,Gao:2021sml}. The reason seems to be obvious. While the electromagnetic form factors can be straightforwardly measured in electron scattering experiments,  there is no known way to directly measure the GFFs due to the weakness of the gravitational interaction. Indeed, if one exchanges a graviton in generic $2\to 2$ scattering, just like the  photon exchange in electron scattering, the cross section behaves as 
\beq
\frac{d\sigma}{dt} \sim G_N^2 \frac{s^2}{t^2}, \label{naive}
\eeq
where $s,t$ are the usual Mandelstam variables and $G_N$ is the Newton constant. Since $G_N\sim 1/M_P^2$ with $M_P\sim 10^{19}$ GeV being the Planck energy, there is no hope for any terrestrial experiment to be sensitive to this enormous suppression, and the cross section will be in any case  completely overwhelmed by  electromagnetic and hadronic backgrounds.  There are, however, indirect ways to access the GFFs. This is based on the idea that two spin-1 particles (like two photons \cite{Teryaev:2005uj,Kumano:2017lhr} or two gluons \cite{Frankfurt:2002ka,Boussarie:2020vmu,Hatta:2021can,Guo:2021ibg}) can mimic the exchange of a spin-2 particle. Another approach is to resort to  holographic QCD models and literally exchange gravitons in higher dimensions \cite{Hatta:2018ina,Mamo:2019mka,Liu:2022zsa}. The first experimental extractions of the GFFs of the proton along these lines have been reported recently \cite{Burkert:2018bqq,Duran:2022xag,GlueX:2023pev}. However, at the moment the precision of such extractions falls short of what is typically  achieved in the determination of the electromagnetic form factors.  Moreover, it is  challenging to systematically improve the precision mainly due to theoretical reasons \cite{Kumericki:2019ddg,Dutrieux:2021nlz,Sun:2021pyw}, except possibly in certain special kinematics \cite{Hatta:2021can,Guo:2021ibg}. Additionally, it has not been demonstrated whether these approaches are practical for the  extraction of the GFF of atomic  nuclei  (see, however,  \cite{Wang:2023uek}). 

In this paper, I explore the possibility of {\it directly} accessing  the GFFs of the proton and nuclei in elastic lepton-proton scattering in the TeV energy regime. The idea is to integrate the measurement of the GFFs with searches for  massive gravitons present in certain scenarios of beyond the Standard Model (BSM) physics and General Relativity. In such scenarios, the effective Planck energy may be in the TeV region, and therefore the suppression  (\ref{naive}) could be compensated if the energy $s$ is sufficiently high. I point out that, in certain regions of the parameter space currently not excluded by the collider data, the massive graviton exchange contribution can be a non-negligible correction to the leading, photon exchange contribution.  Ideal experiments to test this idea are the Large Hadron electron Collider (LHeC) \cite{LHeC:2020van} at CERN and the  muon-ion collider (`MuIC') \cite{Acosta:2021qpx,Acosta:2022ejc} recently proposed as a future upgrade of the Electron-Ion Collider (EIC) \cite{AbdulKhalek:2021gbh} at Brookhaven National Laboratory. \\

 I introduce a massive spin-2 field   $h_{\mu\nu}$ (`graviton') with mass $m$   
 described by the Fierz-Pauli theory \cite{Fierz:1939ix} which is the standard theory of linearized massive gravity. I do not assume any specific  scenario regarding the origin of the field. It may be an effective low energy theory of massive gravity with a nonlinear completion (see e.g., \cite{Schmidt-May:2015vnx}), or alternatively, it could be regarded  simply as a new   spin-2 particle. The field  interacts with  
the Standard Model particles through their energy momentum tensor $T^{\mu\nu}$  as 
\beq
\delta{\cal L} = \kappa h_{\mu\nu}T^{\mu\nu}. \label{model}
\eeq 
I naturally assume universal coupling, that is, $\kappa$ is the same for all the fields in the SM, although this condition can be relaxed in a model-dependent way.\footnote{Such an extension may be interesting to consider because, if the graviton couples to quarks and gluons  differently $\kappa T^{\mu\nu}_{\rm QCD}\to \kappa_q T^{\mu\nu}_q+\kappa_g T^{\mu\nu}_g$, one can  probe the quark and gluon parts of the GFFs.} 

Historically, the potential existence of a new spin-2 particle has attracted significant attention since it can yield observable consequences such as the deviation from the inverse-square law. Over the past several decades,  
the parameters $(m,\kappa)$ have been severely constrained by laboratory experiments and astrophysical  observations  which point to very small values $\kappa < 10^{-20}\sim 10^{-10}$ GeV$^{-1}$ \cite{Talmadge:1988qz,Adelberger:2003zx,Cembranos:2017vgi}.  However, if $m$ is large, say $m\gg 100$ MeV, constraints mainly come from collider experiments in the TeV energy region which are not as restrictive. According to  \cite{Tang:2012pv,Cembranos:2021vdv,dEnterria:2023npy}, 
 values of $\kappa$ as large as $10^{-4} \sim 10^{-3}$ GeV$^{-1}$ are not excluded. (Ref.~\cite{dEnterria:2023npy} constrains the graviton-photon coupling. The result is roughly consistent with  \cite{Cembranos:2021vdv} under the assumption of universal coupling.)     
 I will focus on this region of the parameter space.

The simplest process that is sensitive to the interaction between $h_{\mu\nu}$ and the nucleon is high energy elastic electron-proton scattering $l+p\to l'+p'$.\footnote{The neutron cannot  be accelerated to high energy. But the neutron GFFs are identical to the proton GFFs up to small isospin-breaking effects.}  The scattering amplitude is dominated by the one-photon exchange
and can be parameterized by the proton's electromagnetic form factors $F_{1,2}(t)$ as 
\beq
i{\cal A}_\gamma = e^2 \bar{u}(l')\gamma^{\mu} u(l)\frac{-ig_{\mu\nu}}{t} \bar{u}(p')\left[\gamma^\nu F_1(t)+\frac{i\sigma^{\nu\lambda}\Delta_\lambda}{2M} F_2(t)\right]u(p),
\eeq
where $\Delta^\mu=p'^\mu-p^\mu = l^\mu-l'^\mu$, $\Delta^2=t$ and $M$ is the proton mass. The lepton mass is neglected. 
The dominant source of corrections to this formula is the QED radiative effect  which can be sizable but is  well understood \cite{Akushevich:2015toa,Ye:2017gyb}. The weak interaction contribution is suppressed by the $Z$-boson mass $|t|/M_Z^2 \ll 1$. (The form factors are usually measured up to $|t|\lesssim 10$ GeV$^{2}$)   However, at very high center-of-mass energy of  order $s=(l+p)^2\sim (1\, {\rm TeV})^2$, there will be another source of  corrections from the massive graviton exchange. The corresponding amplitude reads 
\beq
i{\cal A}_G= -\kappa^2 \bar{u}(l')L^{(\mu} \gamma^{\nu)} u(l) G_{\mu\nu\alpha\beta}(\Delta)\bar{u}(p')\left[\gamma^{(\alpha} P^{\beta)} A(t) + \frac{P^{(\alpha}i\sigma^{\beta)\lambda}\Delta_\lambda }{2M}B(t)+\frac{D(t)}{4M}(\Delta^\alpha\Delta^\beta -g^{\alpha\beta}t)\right]u(p), 
\label{ag}
\eeq
where $L^\mu=\frac{l^\mu+l'^\mu}{2}$, $P^\mu=\frac{p^\mu+p'^\mu}{2}$  
 and $G$ is the graviton propagator  
\beq
G_{\mu\nu\alpha\beta}(\Delta)=\frac{i}{t-m^2}\left( \frac{1}{2}(P_{\mu\alpha}P_{\beta\nu}+P_{\mu\beta}P_{\nu\alpha})-\frac{1}{3}P_{\mu\nu}P_{\alpha\beta}\right), \qquad P_{\mu\nu}=g_{\mu\nu}-\frac{\Delta_\mu \Delta_\nu}{m^2}.
\eeq
The graviton-proton coupling is parameterized by the gravitational form factors $A(t),B(t),D(t)$.  From momentum and angular momentum conservation, $A(0)=1$ and $B(0)=0$, but $D(0)$ is not constrained. 

The photon and graviton amplitudes interfere at the amplitude level (see a related discussion in \cite{Zhang:2023nsk}). The unpolarized differential cross section is  given by 
\beq
\frac{d\sigma}{dt}&=&\frac{1}{16\pi s^2}\frac{1}{4}\sum_{\rm spins}|{\cal A}_\gamma+{\cal A}_g|^2 \nn
 &=&  \frac{4\pi \alpha_{em}^2}{t^2} \biggl\{ \left(1+ \frac{t-2M^2}{s} +\frac{M^4}{s^2}\right)\left(F_1^2(t)-\frac{tF_2^2(t)}{4M^2}\right)  +  \frac{t^2}{2s^2}(F_1(t)+F_2(t))^2 \biggr\} \label{dsigma} \\
&&  + \frac{\alpha_{em}\kappa^2s }{t(t-m^2)}\left\{\left(1+ \frac{3(t-2M^2)}{2s}\right)\left(A(t)F_1(t)-\frac{tB(t)F_2(t)}{4M^2}\right) 
+{\cal O}(s^{-2})\right\} 
+{\cal O}(\kappa^4).
\nonumber
\eeq
The first line is the usual Rosenbluth formula in disguise. The ${\cal O}(\kappa^2)$ terms come from the interference effect where the overall factor of $s$ is the hallmark of the spin-2  graviton exchange. Note that the $D(t)$ form factor drops out completely because $G_{\mu\nu\alpha\beta}(\Delta^\alpha \Delta^\beta-g^{\alpha\beta}t)\propto \Delta_\mu \Delta_\nu$ and the leptonic tensor is conserved. This is so even after including the lepton mass or  QED radiative corrections.   The measurement of $D(t)$ has to be done by other means \cite{Teryaev:2005uj,Hatta:2018ina}.

\begin{figure}
    \centering
    \includegraphics[scale=0.5]{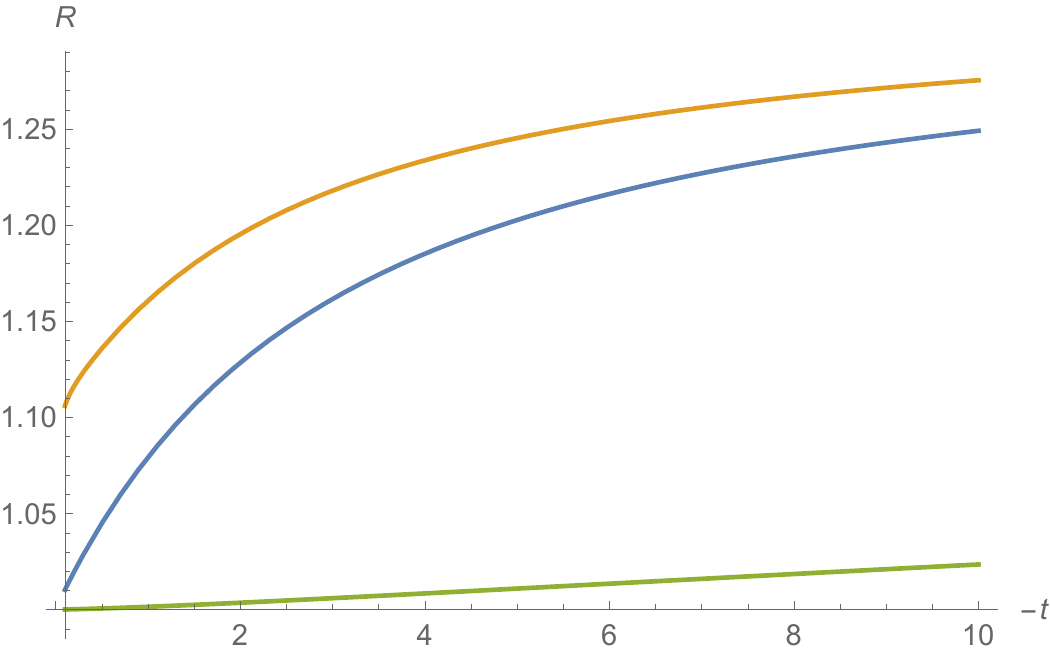}
    \caption{ The ratio (\ref{ratio}) as a function of $-t$ (in GeV$^2$) with $m=0.1$ GeV (top), $m=1$ GeV (middle) and $m=10$ GeV (bottom).  }
    \label{1}
\end{figure}

The magnitude of the interference term relative to the leading term is roughly characterized by the number 
\beq
\frac{\kappa^2 s}{4\pi\alpha_{em}} \sim 10\kappa^2 s . \label{number}
\eeq
If $\kappa \sim 10^{-4}$ GeV, then (\ref{number}) is about 0.1\% when $\sqrt{s}\sim 100$ GeV (EIC) and 10\% when $\sqrt{s}\sim 1$ TeV (LHeC, MuIC).  
To be more quantitative, in Fig.~\ref{1} I plot the ratio 
\beq
R(t)\equiv  \frac{d\sigma/dt|_{\kappa \neq 0}}{d\sigma/dt|_{\kappa=0}}, \label{ratio}
\eeq
for  $\kappa=10^{-4}$ GeV$^{-1}$, $\sqrt{s}=1$ TeV and three different values of the graviton mass $m=0.1, 1,10$ GeV.   I  used  the dipole model for the form factors 
\beq
G_E(t)=\frac{G_M(t)}{\mu_p} =  \frac{1}{\left(1-\frac{t}{0.71{\rm GeV}^2}\right)^2}, \qquad A(t)=\frac{1}{\left(1-\frac{t}{M^2_A }\right)^2}, 
\eeq
where $G_E(t)=F_1(t)+\frac{t}{4M^2}F_2(t)$ and $G_M(t)=F_1(t)+F_2(t)$ are the Sachs form factors and $\mu_p=2.79$ is the proton magnetic moment in units of the nuclear magneton. More realistic estimates based on the state-of-the-art electromagnetic form factors, including QED radiative corrections, are certainly possible   \cite{Akushevich:2015toa,Ye:2017gyb}. As for the $A(t)$ form factor, the dipole form is motivated by the power-counting argument \cite{Tanaka:2018wea,Tong:2021ctu} and I set $M_A\approx 1.4$ GeV from a recent lattice calculation \cite{Hackett:2023rif}. The $B(t)$ form factor has been neglected for simplicity, 
but it can be restored given a suitable parameterization.   

Fig.~\ref{1} shows that an upward  deviation from the QED expectation serves as a signal of the graviton exchange and at the same time carries nontrivial information about the $A(t)$ form factor. Upward, because the electromagnetic and gravitational forces between an electron and a proton are both attractive. The increase of the ratio $R(t)$ with $|t|$ can be attributed to the difference in the slope parameter of the form factors
\beq
F_1(t) \sim \mu_p\frac{0.71^2 {\rm GeV}^4}{t^2}, \qquad   A(t) \sim \frac{M_A^4}{t^2}
\eeq
resulting in an  enhancement factor  $A(t)/F_1(t)\sim 1.4^4/(0.71^2\mu_p) \approx 2.7$ at large-$|t|$. While this number is model-dependent, that $A(t)$ is larger than $F_1(t)$ at large-$|t|$ is expected on general grounds. The $A(t)$ form factor is dominated by the $2^{++}$ glueballs \cite{Mamo:2019mka,Fujita:2022jus} which are heavy, and therefore it decays more slowly with $|t|$ than  the electromagnetic form factors.

It is worthwhile to comment  that, 
according to the analyses in \cite{Cembranos:2021vdv,dEnterria:2023npy}, the largest coupling currently allowable is  $\kappa \sim 10^{-3}$ GeV$^{-1}$, albeit in a  narrow window of $m$. This leaves open the possibility that the signal could be seen already at the EIC top energy  $\sqrt{s}\approx 140$ GeV. The large leverage $-t \sim 40$ GeV$^2$ achievable at the EIC (see Section 7.2.1 of \cite{AbdulKhalek:2021gbh}) may also help.

The generalization to  nuclear targets is straightforward. For spin-$\frac{1}{2}$ nuclei, the cross section is given by the same formula (\ref{dsigma}) except for the substitutions  $\alpha_{em}\to Z\alpha_{em}$ and $s\to s_{eA}=A s_{ep}$ where  $Z,A$ are the charge and mass numbers, respectively, and $\sqrt{s_{ep}}$ is the center-of-mass energy per nucleon.  The interference term gets an extra relative  enhancement factor $A/Z\gtrsim 2$. For spin-0 nuclei, the $B(t)$ form factor is absent. For nuclei with spin larger than $\frac{1}{2}$, there are more GFFs characterizing the nuclear shape and polarization, and some of them survive at leading order in $s$.  For example, the deuteron (spin-1) has six independent GFFs  including \cite{Holstein:2006ud}
\beq
\langle p'|T^{\mu\nu}|p\rangle \sim \frac{2}{M^2}P^\mu P^\nu \epsilon\cdot \Delta \, \epsilon'^*\cdot \Delta F_5(t), \label{f5} 
\eeq
where $\epsilon^\mu$ is the polarization vector. 
(\ref{f5}) contributes to the $\cos 2\phi$ azimuthal angle  correlation between $\vec{\Delta}$ and the linearly polarized vector $\vec{\epsilon}$.  
Note however that  the nuclear electromagnetic form factors decay faster with increasing $|t|$, and the same is true for the nuclear GFFs \cite{GarciaMartin-Caro:2023klo,He:2023ogg}. In practice, this will limit measurements to smaller-$t$ regions where the ratio $R$ is also smaller.

Finally, it should be cautioned that the above argument exploits the fact that the $t$-channel graviton propagator $1/(t-m^2)$ is not suppressed by a large (TeV) scale  for the values of $t$ and $m$ considered in this paper.  This is not always be the case.  In  extra dimension models \cite{Arkani-Hamed:1998sfv}, the coupling with each Kaluza-Klein (KK) graviton is suppressed by the Planck scale $\kappa \sim 1/M_P$. However, there are infinitely many of them, and after summing over the exchanges of all the KK gravitons one obtains an effective local interaction   
\beq
\delta {\cal L} =\frac{4}{M_{TT}^4}T^{\mu\nu}T_{\mu\nu},
\eeq
with $M_{TT}\sim {\cal O}({\rm TeV}) \ll M_P$, irrespective of $s,t$ or $u$-channel \cite{Han:1998sg} 
(up to a logarithmic enhancement $\ln\frac{M^2_{TT}}{|t|}$ in the case of  $n=2$ extra dimensions). The cross section is then given by (\ref{dsigma}) with  the replacement  $\frac{\kappa^2}{t-m^2}\to \frac{8}{M_{TT}^4}$.  The current 95\% CL lower bound from the LHC data is $M_{TT}\gtrsim 6\sim 9$ TeV \cite{ParticleDataGroup:2022pth}. This means that the interference term in (\ref{dsigma}) is relatively suppressed at least by a factor $10^{-4}$ even when $\sqrt{s}=1$ TeV. A similar conclusion may be drawn for warped extra dimension models  \cite{Randall:1999ee} where $m \sim {\cal O}(1\, {\rm  TeV})$.   Therefore, in extra dimension scenarios  the present approach requires a very precise measurement of the elastic cross section or even higher center-of-mass energies.

In conclusion, I have proposed a novel  method to  directly access the gravitational form factors at future lepton-ion colliders such as the LHeC and MuIC, and possibly also at the EIC. Conversely, with the help of a reasonable model for the GFFs,  the measurements can constrain the parameters of massive gravity models. The proposal is unique in that, unlike in most BSM physics searches, the `QCD part' is not an unexciting background but can reveal some of the fundamental aspects of the nucleon/nuclear structure that the EIC is built for. Therefore, it can benefit from an interdisciplinary effort by the hadron and BSM physics communities.    

\begin{acknowledgements}
I thank Hooman Davoudiasl for helpful discussion. 
This work is supported by the U.S. Department of Energy under Contract No. DE-SC0012704, and also by  Laboratory Directed Research and Development (LDRD) funds from Brookhaven Science Associates. 
\end{acknowledgements}

\bibliography{Bibliography}
\end{document}